\documentclass[]{aa}
\usepackage{graphics,latexsym}     
\begin{document}
\title{Parameterised models for the lensing cluster Abell 1689}
\author{L.J. King$^{1,2}$, D.I. Clowe$^{2}$, \& P. Schneider$^{2,1}$ }
\institute{1: Max-Planck-Institut f{\"u}r Astrophysik,
Karl-Schwarzschild Str 1, Garching bei M{\"u}nchen, Germany\\
2: Institut f{\"u}r Astrophysik und Extraterrestrische
Forschung der Universit{\"a}t Bonn, Auf dem H{\"u}gel 71, D-53121 Bonn,
Germany\\
}
\date{}
\authorrunning{King, Clowe \& Schneider}
\abstract{
Here we apply a recently developed maximum likelihood method for 
determining best-fit parameterised lens models to observations of the 
rich lensing cluster Abell 1689. The observations that we use were taken 
with the ESO/MPG Wide Field Imager. The wide field-of-view enables us
to use the weakly lensed images of faint background objects on an
unsurpassed range of scales, $0.12\,h^{-1}\,{\rm Mpc} < R < 1.8\,h^{-1}\,{\rm Mpc}$ from the
cluster centre, to determine best-fit models for the 1-parameter singular 
isothermal sphere (SIS), 2-parameter general power-law and NFW
models, and 3-parameter singular isothermal ellipsoid (SIE). 
The best-fit SIS has an Einstein radius $\theta_{\rm E}=0\arcminf 37$
$(0.043\,h^{-1}\,{\rm Mpc})$ i.e. a velocity dispersion $\sigma_{\rm 1D}=1028^{+35}_{-42}\,{\rm km\,s^{-1}}$ in an $\Omega=1.0$, $\Lambda=0.0$ cosmology. For the best-fit NFW
profile, the virial radius $r_{200}=1.14\,h^{-1}\,{\rm Mpc}$ and the 
concentration parameter $c=4.7$, giving a virial mass $M_{200}=5.7\times
10^{14}\,h^{-1}\,{\rm M}_{\odot}$. At $q=0.88$, the slope of the best-fit 
power-law model is slightly flatter than isothermal ($q=1.0$), indicating that the galaxies most
important in the fitting procedure lie inside the scale radius
$r_{\rm s}$. By fitting an SIE, the deviation of the projected mass
distribution from circular symmetry is evident, with a best-fit axial
ratio $f=0.74$. 
}

\def\A{{\cal A}}
\def\eck#1{\left\lbrack #1 \right\rbrack}
\def\eckk#1{\bigl[ #1 \bigr]}
\def\rund#1{\left( #1 \right)}
\def\abs#1{\left\vert #1 \right\vert}
\def\wave#1{\left\lbrace #1 \right\rbrace}
\def\ave#1{\left\langle #1 \right\rangle}
\def\arcsecf {\hbox{$.\!\!^{\prime\prime}$}}
\def\arcminf {\hbox{$.\!\!^{\prime}$}}
\def\bet#1{\left\vert #1 \right\vert}
\def\vp{\varphi}
\def\vt{{\vartheta}}
\def\d{{\rm d}}
\def\eps{{\epsilon}}
\def\vc{\vec} 
\def\s{{\rm d}}
\def\s{{\rm s}}
\def\t{{\rm t}}
\def\E{{\rm E}}
\def\L{{\cal L}}
\def\q{{\rm \i}}

{\catcode`\@=11
\gdef\SchlangeUnter#1#2{\lower2pt\vbox{\baselineskip 0pt \lineskip0pt
  \ialign{$\m@th#1\hfil##\hfil$\crcr#2\crcr\sim\crcr}}}
}
\def\gtrsim{\mathrel{\mathpalette\SchlangeUnter>}}
\def\lesssim{\mathrel{\mathpalette\SchlangeUnter<}}      
\maketitle
\section{Introduction}
The gravitational field of a cluster acting as a weak lens distorts the
shapes of background galaxies (shear effect), and changes their
observed number density (magnification effect); see Mellier (1999) and
Bartelmann \& Schneider (2001) for recent reviews. These
imprints on the background galaxy population can be used to constrain
the projected mass distribution of the cluster.
Parameterised cluster models are essential when statistical comparisons of
clusters are to be made, and their error properties are well
understood in contrast to those of non-parametric mass maps. 

In Schneider et al. (2000; hereafter SKE) we developed 
likelihood techniques to obtain parameterised models from weak lensing 
data, and quantified the
accuracy with which parameters can be recovered using the
magnification and/or shear information. This
work concentrated on power-law models, and 
was extended in King \& Schneider (2001; hereafter KS) to 
encompass NFW  (Navarro et al. 1996; 1997) and ellipsoidal models
(Kormann et al. 1994), also investigating the effect of uncertainty in 
the redshifts of potential source galaxies. In King et al. (2001; hereafter KSS) we considered the
influence of substructure, concluding that the method is robust
against the presence of substructure at a level consistent with 
N-body simulations. Gray et al. (2000) applied the magnification
maximum likelihood method to near-infrared CIRSI observations of the
cluster Abell 2219, and extended it to include the effect of 
incompleteness.

In this paper we obtain parameterised models of a
lensing cluster by applying the shear maximum likelihood method. 
We use WFI observations of the rich $z_{\d}=0.18$ lensing cluster Abell 1689, 
the observational details of which are presented in Clowe \& Schneider
(2001). 

The families of lens models considered are the
singular isothermal sphere (SIS), the singular isothermal ellipsoid (SIE), 
a general power-law model and the NFW profile. The SIS model is
frequently used as the basis model for
lensing studies, and the introduction of an ellipticity, yielding the SIE
model, allows us to investigate non-spherically symmetric dark matter 
distributions. 
The motivation for considering the NFW profile is
that it is a good description of the radial density profiles of
virialised dark matter halos formed in cosmological simulations of
hierarchical clustering. We also investigate a general single slope 
power-law model, since in KSS we found that lensing simulations 
through an N-body cluster were well fit by this model.

The structure of this work is as follows: in Sect. 2 we
summarise how the shear information is incorporated into a likelihood 
analysis, and outline the relevant observational parameters. 
Our results are presented in Sect. 3, and we finish with a
discussion and conclusions in Sect. 4.

\section{An outline of the shear method and observational details}
The notation and basic relationships used in this paper are as in KS,
and the reader is referred to SKE for details of the maximum
likelihood treatment. Throughout, standard lensing 
notation is used (e.g. Schneider, Ehlers \& Falco 1992; Bartelmann \&
Schneider 2001). 
 
\subsection{The shear method}
The basis of using the distorted images of background galaxies to 
constrain the cluster model is that the expectation value for the 
lensed ellipticity $\ave{\eps}=g$, the reduced shear, in the 
non-critical regime (e.g. Schramm \& Kayser 1995), and that 
$\ave{\eps}=1/g^{*}$ in the critical regime (Seitz \& Schneider 1997).

A parameterised model specifies the
dependence of $g(\vc\vt)$ on position $\vc\vt$ for a profile family; 
these are given in Sect. \ref{profiles} below, for the families
considered. Minimising the shear log-likelihood function (see SKE for details)
\begin{equation}
\ell_{\gamma}=-\sum_{i=1}^{N_{\gamma}}{\rm\ln\;}p_{\epsilon}(\epsilon_{i}|g(\vc\vt_{i}))\;
\label{liksh}
\end{equation}                        
over the $N_{\gamma}$ galaxy images gives $\pi_{\rm max}$, the
parameters most consistent with the probability distribution of lensed 
ellipticities $p_{\epsilon}\left(\epsilon|g(\vc\vt)\right)$. 
The trial parameters en route to the best-fit will be referred to by 
$\pi$. Noise in this method arises from the intrinsic dispersion in the galaxy
ellipticity distribution, $\sigma_{\eps^{\s}}$, and observational effects.

\subsection{Observational parameters}

The catalogue of lensed galaxies in the cluster A1689 containing positions $\vc\vt_{i}$
and lensed ellipticities  $\epsilon_{i}$ is described in Clowe \&
Schneider (2001). Objects therein were selected to have $23<m_{R}<25.5$, and a 
signal-to-noise greater than 9. The data field used in the method has 
inner and outer radii of $\theta_{\rm in}=1\arcminf 0$ (0.12$\,h^{-1}\,$Mpc) and 
$\theta_{\rm out}=15\arcminf 4$ (1.8$\,h^{-1}\,$Mpc) respectively. 
Within this aperture, there are $N_{\gamma}=19400$ images for which an ellipticity $\eps_{i}$ can be measured, and
can therefore be used for the shear method. 
These galaxies have a Gaussian probability distribution 
with 2-D dispersion $\sigma_{\epsilon}=0.424$; since $|g| \lesssim 0.2$,
correction to the unlensed dispersion is not necessary.

The redshift distribution of the source galaxy population can safely
be neglected, since the lens is at a fairly low redshift 
$(z_{\rm d}=0.18)$  (see for example Bartelmann \& Schneider 2001). 
This amounts to approximating the background galaxies to be located at
a redshift
corresponding to the mean value of their lensing effectiveness
parameter $w(z)$. If the cluster was at a higher redshift, $\gtrsim 0.25$ say, then the redshift
distribution of the galaxies would become important and this sheet
approximation is no longer robust. We take the background sources to
be at $z=1.0$ 
unless otherwise stated. This is motivated by Clowe \& Schneider
(2001), who used the HDF-South photometric redshift catalogue from
Fontana et al. (1999) and found a mean galaxy redshift of 1.15 after
applying the same magnitude cuts as above. Since this value is
affected by Poisson noise and cosmic variance, being measured from
only 48 galaxies, we consider the
influence of the assumed source redshift in Sect. \ref{cosmo}.
Unless otherwise stated, we give the results for an $\Omega=1.0$,
$\Lambda=0.0$ cosmology, for a Hubble constant $H_{0}=100\,h\,{\rm km\,s}^{-1}\,{\rm Mpc}^{-1}$. At $z_{\rm d}=0.18$, an angular scale of
$1\arcminf 0$ corresponds to a linear scale of 0.12$\,h^{-1}\,{\rm Mpc}$.

\subsection{Profiles \label{profiles}}

\subsubsection{The isothermal profile}

The singular isothermal profile, or SIS, is characterised by a
velocity dispersion $\sigma_{\rm 1D}$ that is constant with radius. It is
parameterised by an Einstein radius 
\begin{equation}
\theta_{\rm E}= \frac{4\pi \sigma_{\rm 1D}^{2}}{c^{2}}\frac{D_{\d\s}}{D_{\s}}
\label{the}
\end{equation}
which marks the boundary between the strong and weak lensing
regimes. Note that for a given $\sigma_{\rm 1D}$, the Einstein radius depends on
$\Omega$ and $\Lambda$ through the angular diameter distances, but
that it is independent of $H_{0}$. The surface mass density is given by
\begin{equation}
\kappa(\theta)=0.5\left(\frac{\theta}{\theta_{\rm E}}\right)^{-1}\;.
\end{equation}
The mean surface mass density inside $\theta$ is 
$\bar\kappa(\theta)=2\kappa(\theta)$, and further $|\gamma|=\kappa$.

\subsubsection{The singular isothermal ellipsoidal profile}

The SIE profile is characterised by a velocity 
dispersion that is
constant with radius; it is parameterised by an equivalent 
angular Einstein radius $\theta_{\rm E}$, an axial ratio $0<f\le 1$
and an orientation $\alpha$.
Let the polar coordinates in the lens plane be 
${\vc\vt} = (\theta{\rm cos}\phi,\theta{\rm sin}\phi)$. The
equivalent angular Einstein radius for a lens with velocity dispersion
$\sigma_{\rm 1D}$ is given by Eq.\ref{the}. The dimensionless surface mass density is given by
\begin{eqnarray}
\kappa(\theta,\phi)&=&\frac{\sqrt{f}}{2b}\nonumber\\
b&=&\frac{\theta}{{\sqrt 2}\,\theta_{\rm E}}\left(\left(1+f^{2}\right)+\left(1-f^{2}\right){\rm cos}[2(\phi-\alpha)]\right)^{\frac{1}{2}}\; 
\label{pop}
\end{eqnarray}
and the magnification is
\begin{equation}
\mu(\theta,\phi)=\frac{1}{1-2\kappa(\theta,\phi)}\;.
\end{equation}
The components of the shear are
\begin{equation}
\gamma_{1}=-\kappa~{\rm cos}(2\phi);~~~~~\gamma_{2}=-\kappa~{\rm sin}(2\phi)\;.
\label{gammy}
\end{equation}

\subsubsection{The power-law profile}

The power-law profile can be 
characterised by an Einstein radius $\theta_{\rm E}$ and a slope $q$, such that
\begin{equation}
\kappa(\theta)=\frac{2-q}{2}\left(\frac{\theta}{\theta_{\rm E}}\right)^{-q}\;,
\end{equation}
and
\begin{equation}
\bar\kappa(\theta)=\left(\frac{\theta}{\theta_{\rm E}}\right)^{-q}\;.
\end{equation} 
The shear is
\begin{equation}
\gamma(\theta)=\frac{q}{2}\left(\frac{\theta}{\theta_{\rm E}}\right)^{-q}\;.
\end{equation}
In KSS we found that the power-law profile can provide almost as good
a fit as the NFW profile, depending on the size of the 
cluster's scale radius $r_{\rm s}$ relative to the available data field.

\subsubsection{The NFW profile}

The properties of the NFW profile in the context of gravitational
lensing have been discussed by several authors including Bartelmann (1996) 
and Wright \& Brainerd (2000).
We can parameterise this profile with a virial radius $r_{200}$, and a 
dimensionless concentration parameter $c$, which are related through a
scale radius $r_{\rm s}=r_{200}/c$. Inside $r_{200}$, the mass density of the
halo equals 200$\rho_{\rm c}$, where $\rho_{\rm c}=\frac{3H^{2}(z)}{8\pi G}$ is the critical density of the Universe at the redshift of the halo. The 
characteristic overdensity of the halo, $\delta_{\rm c}$, is related to 
$c$ through
\begin{equation}
\delta_{\rm c} = \frac{200}{3}\frac{c^{3}}{{\rm ln}(1+c)+c/(1+c)}\;.
\end{equation}
Then the density profile is
\begin{equation}
\rho(r) = \frac{\delta_{\rm c}\rho_{\rm c}}{(r/r_{\rm s})(1+r/r_{\rm s})^{2}}\;,
\end{equation}
which is shallower than isothermal ($r^{-2}$ in 3-D) near the halo centre and 
steeper than isothermal for $r\gtrsim r_{\rm s}$. For details of
quantities required in the lensing analysis, see KS.

\section{Results}

\subsection{The best-fit parameterised models}
The shear method was applied to the catalogue of background objects,
finding the best-fit model parameters, for each of the model
families described above. 

For the 1-parameter SIS, the 
best-fit $\theta_{\E}$=$0\arcminf 37$; this corresponds to a linear size of
0.043$\,h^{-1}\,{\rm Mpc}$ at the lens redshift. The values
of $2\Delta\ell\equiv 2\left[\ell(\pi)-\ell(\pi_{\rm max})\right]$
were calculated from the data in order to obtain confidence levels for the fit.
The upper (lower) 1-$\sigma$ limits are $0\arcminf 395~(0\arcminf 34)$
and the upper (lower) 3-$\sigma$ limits are $0\arcminf 44~(0\arcminf
29)$. This translates into $\sigma_{\rm 1D}=1028^{+35}_{-42}{\rm
km\,s}^{-1}$ (1-$\sigma$ errors), for $\Omega=1.0$ and $\Lambda=0.0$.
In an $\Omega=0.3$, $\Lambda=0.7$ Universe we obtain 
$\sigma_{\rm 1D}=998^{+33}_{-42}{\rm km\,s}^{-1}$ (1-$\sigma$ errors); this 
is within the 1-$\sigma$ error of the result given in Clowe \& Schneider 
(2001) for the same choice of cosmology and background source
redshift, calculated with a direct $\chi^{2}$ fit to the shear profile.   

In Table \ref{res}, we present the values of
$\pi_{\rm max}$ most consistent with the data for the 2-parameter 
power-law and NFW profiles, and for the 3-parameter SIE profile.
Comparing the 2-parameter families, the overall best-fit model comes 
from the NFW family but it is only marginally preferred over the 
power-law model. For 2 degrees of freedom, the values of 
$2\Delta\ell\equiv 2
\left[
\ell\left(\pi_{\rm max}\left({\rm power-law}\right)\right)-
\ell\left(\pi_{\rm max}\left({\rm NFW}\right)\right)
\right]$ show that the best-fit models are only distinguished at about 
80\% confidence. 

In order to fit the SIE profile, we adopted a complex representation
for the ellipticity of the lensing cluster:
\begin{equation}
\epsilon_{\rm A1689}=\frac{1-f}{1+f}{\rm e}^{2{\rm i}\alpha}\;
\end{equation}
and minimised $\ell$ in $\epsilon_{1}-\epsilon_{2}-\theta_{\rm E}$
parameter space. Fitting this profile shows that the data require an 
asphericity in the lensing mass distribution, which is also suggested both 
by the luminous mass distribution and by non-parametric mass reconstruction. 

\begin{table}
\caption{The best-fit parameters $\pi_{\rm max}$ for the SIE,
power-law (POW) and NFW profiles are presented in the middle
column. The final column shows $2\Delta\ell\equiv 2\left[\ell\left(\pi_{\rm max}\left({\rm Model}\right)\right)-\ell\left(\pi_{\rm max}\left({{\rm Model}_{\rm NFW}}\right)\right)\right]$.}
\begin{tabular}{|l|l|l|}
\hline
Model&Best-fit parameters~~$\pi_{\rm max}$&2$\Delta\ell$\\
\hline
SIE&$\theta_{\rm E}=0\arcminf 37$;~~~$f=0.74$; $\alpha=-27^{\circ}$&2.5\\
POW&$\theta_{\rm E}=0\arcminf 3$;~~~$q=0.88$&3.2\\
NFW&$r_{200}=1.14\,h^{-1}\,{\rm Mpc}$;~~~$c=4.7$&--\\
\hline
\end{tabular}
\label{res}
\end{table}

\begin{figure}
\resizebox{8cm}{!}{\includegraphics{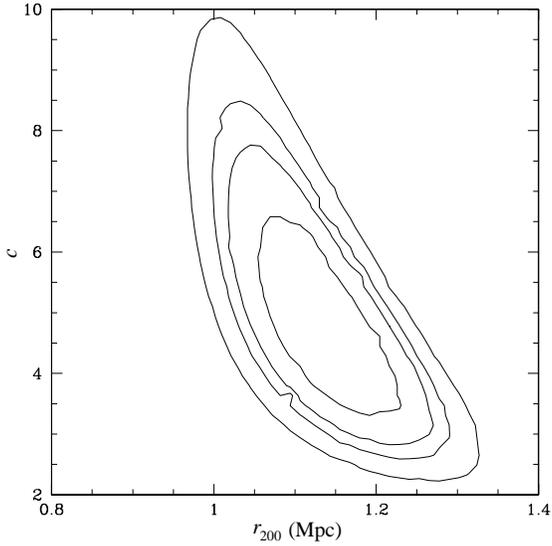}}
\hfill
\caption{
This figure shows the confidence contours for the NFW profile
family. We show contours of constant
$2\Delta\ell=2\left[\ell\right(\pi\left)-\ell\left(\pi_{\rm
max}\right)\right]$. Starting with the innermost contour, the levels are  
plotted at 68.3, 90, 95.4 and 99-\% confidence.}
\label{likes}
\end{figure}

\begin{figure}
\resizebox{8cm}{!}{\includegraphics{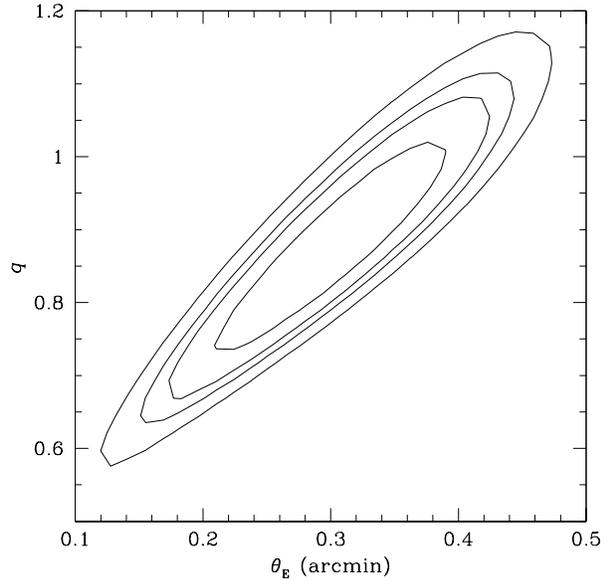}}
\hfill
\caption{
Same as Fig.\ref{likes} for the power-law family.}
\label{likespow}
\end{figure}

For the NFW model in an $\Omega=1.0, \Lambda=0.0$ Universe, confidence
contours $(2\Delta\ell=2\left[\ell\right(\pi\left)-\ell\left(\pi_{\rm
max}\right)\right])$ are shown in Fig.\ref{likes}. Confidence contours 
for the power-law model are shown in Fig.\ref{likespow}.

\begin{figure}
\resizebox{8cm}{!}{\includegraphics{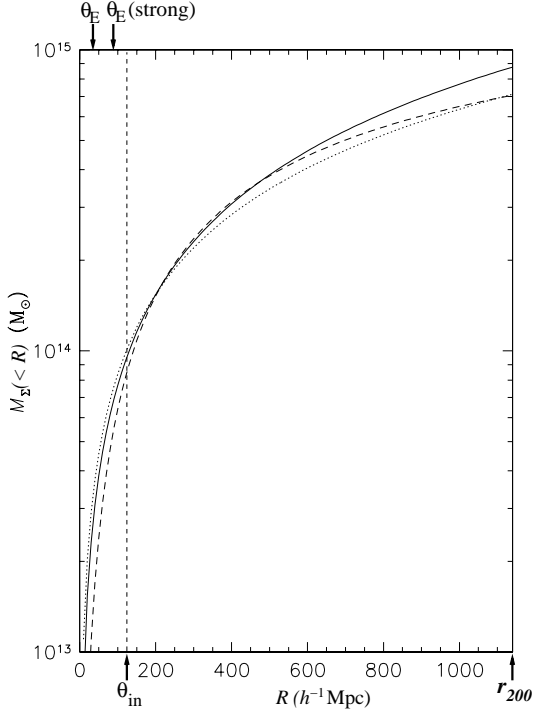}}
\hfill
\caption{
The behaviour of the enclosed surface mass $M_{\Sigma}(<R)$ as a
function of projected distance $R$ from the cluster centre, plotted
out to $r_{200}=1.14\,h^{-1}\,{\rm Mpc}$ for the
best-fit SIS (solid line), POW (dotted line) and NFW (dashed line)
models. The inner fitting radius, $\theta_{\rm in}$ is indicated by a
vertical dashed line (the outer fitting radius, $\theta_{\rm out}$, is 
larger than $r_{200}$). On the upper horizontal axis, the best-fit SIS 
Einstein radius from weak lensing
is denoted by $\theta_{\rm E}$, and the Einstein radius implied by 
the strong lensing arcs by $\theta_{\rm E}$(strong).}
\label{profs}
\end{figure}

Fig.\ref{profs} shows the enclosed projected surface mass
$M_{\Sigma}(<R)$ (in units of ${\rm M}_{\odot}$) for each of the SIS,
POW and NFW models, as a function of projected distance from the
profile centre $R$ (in kpc) plotted out to $R=r_{200}=1.14\,h^{-1}\,{\rm Mpc}$. Note the good agreement between the total masses estimated by fitting each of the models: the maximum difference beween the models is $\sim 20\%$ at the virial radius and smaller for smaller radii.

\subsection{Assessing the goodness of fit}
\begin{figure}
\resizebox{9cm}{!}{\includegraphics{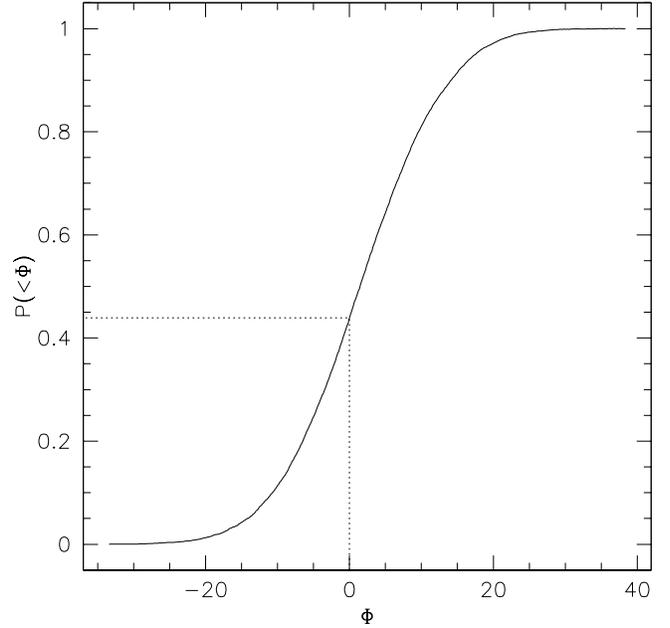}}
\hfill
\caption{Goodness of fit results for the NFW model: the distribution of P$(<\Phi)$ for 
$\Phi:=2\left[\ell_{R}(\pi_{\rm max})-\ell(\pi_{\rm max})\right]$
determined from 10000 randomisations of the background galaxy position
angles.}
\label{siggy}
\end{figure}

Just how good a fit is the best-fit NFW model, $\pi_{\rm max}$? 
Specifying $g(\vc\vt)$ for $\pi_{\rm max}$, the catalogue was
``unlensed" by transforming the observed $\epsilon_{i}$ to intrinsic
$\epsilon_{i}^{\rm s}$ under this lens model. Taking this reference catalogue, the phases of 
$\epsilon_{i}^{\rm s}$ were randomised, while preserving $\vc\vt_{i}$, 
and $|\epsilon_{i}^{\rm s}|$, to obtain a new catalogue. This catalogue
was lensed and the value of 
$\ell_{R}(\pi_{\rm max})$, where the subscript $R$
denotes randomisation, was determined for this realisation. 
This randomisation of the reference catalogue, lensing and 
determination of $\ell_{R}(\pi_{\rm
max})$ was repeated 10000 times, to build up statistics of
$\Phi:=2\left[\ell_{R}(\pi_{\rm max})-\ell(\pi_{\rm max})\right]$; the
cumulative probability distribution P($<\Phi$) is shown in
Fig.\ref{siggy}. The value $\ell(\pi_{\rm max})$ (fit to the real data
set) lies well within the distribution of $\ell_{R}(\pi_{\rm max})$ for the 
random phase data sets, with 43.9 (56.1)\% of realisations being 
more (less) consistent with the model.  

We repeated the process for the power-law model, and in 
this case, 44.3 (55.7)\% of randomised data sets
are more (less) consistent with the model.  

\subsection{The influence of the aperture size}

In order to test the influence of the size of the data field on the
recovered NFW parameters, the value of $\theta_{\rm out}$ was 
decreased from $15\arcminf 4$ and $\pi_{\rm max}$ obtained for these 
catalogues. 
The largest deviation of $\pi_{\rm max}$ for the smaller apertures
compared with the full aperture is at most 10\% until 
$\theta_{\rm out}\sim 8\arcminf 0$. For instance, when 
$\theta_{\rm out}=10\arcminf 0$ the deviation of both $r_{200}$ and
$c$ from their values when $\theta_{\rm out}=15\arcminf 4$ is well
under 1\%. Below about $8\arcminf 0$, the shear signal
becomes noisy, and changing the aperture slightly can have a
significant effect on parameter recovery. 
Another effect of decreasing aperture size is to expand the confidence 
contours of the best-fit model; this has been discussed in detail in 
SKE. 

When $\theta_{\rm out}=15\arcminf 4$ and $\theta_{\rm in}$
is increased from $1\arcminf 0$ to $2\arcminf 0$, the best-fit NFW
model still lies within the original 68.3-\% confidence contour.

\subsection{The impact of background source redshift and
cosmology\label{cosmo}}
Concentrating on the NFW profile to examine the impact of our choice
of background source redshift and cosmology, Table \ref{res2} shows
the values of $\pi_{\rm max}$ obtained with various assumptions.
To facilitate the comparison between the different
scenarios, we also give values for $M_{200}$, the mass contained within a
sphere of radius $r_{200}$ and having mean density $200\rho_{c}$.
Results are shown for an $\Omega=1.0, \Lambda=0.0$ cosmology, and 
for an $\Omega=0.3, \Lambda=0.7$ cosmology with $z=1$. 

If the background sources are at $z=1.25$ rather than at
$z=1.0$, $M_{200}$ would be about 5\% lower, and if the sources were at
$z=0.75$ the mass estimate would be about 10\% higher. 
Working in an $\Omega=0.3, \Lambda=0.7$ cosmology, and keeping
$z=1.0$, gives an increase of about 5\% in the best-fit value of $M_{200}$.
Again, we can compare this result with the value obtained by 
Clowe \& Schneider assuming the same cosmology and $z=1.0$: the values
of $r_{200}$ and $c$ in that work correspond to an $M_{200}$ value
that is 2.5\% higher than that obtained here. 

\begin{table}
\caption{This Table shows $\pi_{\rm max}$ and the corresponding value
of $M_{200}$ for
the NFW profile, assuming different redshifts for the
background sources, $z$. An $\Omega=1.0$, $\Lambda=0.0$ Universe
is denoted by EdS, and an $\Omega=0.3$, $\Lambda=0.7$ Universe by $\Lambda$-flat.}
\begin{tabular}{|l|l|l|l|l|}
\hline
Cosmology&$z$& $r_{200}$&$c$&$M_{200}$\\
&&$(h^{-1}\,{\rm Mpc})$&&$(10^{14}\,h^{-1}\,{\rm M}_{\odot})$\\
\hline
EdS&0.5&1.27&5.2&7.8\\
EdS&0.75&1.18&4.8&6.3\\
EdS&1.0&1.14&4.7&5.7\\
EdS&1.25&1.12&4.6&5.4\\
\hline
$\Lambda$-flat&1.0&1.29&4.8&5.95\\
\hline
\end{tabular}
\label{res2}
\end{table}

\section{Discussion and conclusions}

The mass profile of a cluster can be constrained by using the
weakly lensed images of background galaxies; since lensing does not
depend on the nature of the matter, it is an invaluable means to
study the luminous and dark matter. One technique that
harnasses the infomation contained in the distorted ellipticities of
the background population is the shear likelihood 
method developed in SKE and in KS. In this paper we have applied the
method to WFI observations of the $z_{\d}=0.18$ lensing 
cluster Abell 1689, to obtain best-fit SIS, SIE, NFW and 
power-law models. Up until now, we had rigorously tested our method
on synthetic data sets, generated by lensing through analytic profiles 
and numerically simulated clusters. 

The maximum likelihood method developed in SKE and extended in KS 
differs from standard $\chi^{2}$ fitting [used recently by 
Clowe \& Schneider (2001) for example] in a few respects:
\begin{itemize}
\item{Each galaxy is treated individually rather than being binned.}
\item{The most likely model is found by finding the parameters
that maximise the probabibility of the lensed ellipticity
distribution. In $\chi^{2}$ fitting, the difference between the model
and observed reduced shear is minimised, taking into account the 
observational errors.} 
\item{The unlensed ellipticity distribution is assumed to be a
gaussian in the maximum likelihood method, whereas the lensed ellipticity distribution is assumed to be a gaussian in $\chi^{2}$ fitting.}
\end{itemize}
Since a subset of the models considered here, namely the NFW and SIS
models, were considered by Clowe \& Schneider (2001), we can 
directly compare the results. 

The distorted images of background sources, at projected distances
between $1\arcminf 0$ and $15\arcminf 4$ ($0.12\,h^{-1}\,{\rm Mpc} < R < 1.8\,h^{-1}\,{\rm Mpc}$) from the centre of the
cluster have been used to obtain the best-fit models. In an 
$\Omega=1.0, \Lambda=0.0$ cosmology, the best-fit 1-parameter SIS
model has $\sigma_{\rm 1D}=1028^{+35}_{-42}{\rm km\,s}^{-1}$. 
This velocity dispersion is consistent with that of the X-ray study of
Allen (1998), who investigated a cluster sample and obtained temperature
profiles by applying a deprojection technique to X-ray surface
brightness profiles.     
Dynamical analyses of the cluster indicate that it is not a simple system: Teague, Carter \& Gray (1990) found 
$\sigma_{\rm 1D}=2355{\rm km\,s}^{-1}$, and evidence for three other
groups of galaxies in its foreground. A more recent analysis by Girardi et al. (1997) finds that A1689 is composed of three structures
for which the resultant $\sigma_{\rm 1D} \sim 560{\rm km\,s}^{-1}$ 
is much lower. Girardi et al. (1997) point out that their result, 
along with the foreground groups suggested by Teague et al. (1990),
may indicate the presence of a filamentary structure along the line of
sight. The strong lensing giant arcs imply an Einstein radius 
$\theta_{\rm E}\approx 0\arcminf 75$ (e.g. Miralda-Escud\'{e} \& Babul
(1995)). Assuming the arcs are at $z=3$, the corresponding SIS has 
$\sigma_{\rm 1D}=1360{\rm km\,s}^{-1}$. In fact, $\sigma_{\rm 1D}$ is rather insensitive to $z$, decreasing by less than 2\% even when $z=6$.

As discussed in Clowe \& Schneider (2001), this data set gives lower limits for the mass of the lensing cluster, due to contamination
of the galaxy catalogue with stars and foreground and cluster dwarf 
galaxies. Even with any realistic correction considered, they
find that the weak lensing value can not be reconciled with that
interpreted from strong lensing data. This points to the need for more
spectroscopic and photometric redshift information in both regimes, 
and a thorough combined analysis of the
weak and strong lensing data.

Formally, the best-fit 2-parameter model belongs to the NFW family 
$(r_{200}=1.14\,h^{-1}\,{\rm Mpc}$, $c=4.7)$, but it cannot be
distinguished from the other best-fit models at 
the 3-$\sigma$ level. In order to distinguish profiles at a
satisfactory significance, so testing the CDM paradigm (that
predicts NFW and similar cuspy profiles) stacking the signals from
a number of clusters is required.

The best-fit general power-law has a slope
$q=0.88$, and $\theta_{\rm E}=0\arcminf 3$ ($0.035\,h^{-1}\,{\rm Mpc}$).
One might have expected a steeper slope, since most of the background
galaxies sample the profile at radii outside $r_{\rm s}$, where the slope
is expected to be steeper than isothermal. Our result indicates that
galaxies close to $\theta_{\rm in}$ are important in the determination
of the best-fit model. 

The confidence contours for the NFW result show that 
$r_{200}$ is better constrained than $c$, consistent
with our study of NFW profiles in KS. Thus $M_{200}$, which depends on
$r_{200}$, is also well constrained.
When the power-law is fit, the confidence contours show a smaller 
fractional uncertainty in $q$ than in $\theta_{\rm E}$,
which is in agreement with our finding in SKE. 

We also considered a 3-parameter SIE model; this model reveals that the
projected (dark and luminous) surface mass density is not circularly 
symmetric.
 The appearance of the cluster on optical images, and a
direct non-parametric mass reconstruction also suggest deviation from
circular symmetery. It is interesting that although the SIE is a
3-parameter model, its likelihood is in fact lower than that of the
2-parameter NFW model. This suggests that in order to fit the weak 
lensing data, any deviation from circular symmetry is not as important 
a driver as the deviation from a purely isothermal profile slope.
Another factor coming into play may also be the ``twisting" of the
isodensity contours as a function of distance from the 
centre; this is evident when the fitting aperture is changed, and even
directly on the non-parametric reconstruction.

By randomising the phases of the background sources, we tested that
the NFW model and power-law models are indeed good fits to the real 
data. One point to note is that the galaxy catalogue must be
unlensed before the phases are randomised and the catalogue is
again lensed. If the catalogue is not unlensed, this results in a
significant offset between the likelihood values from these randomisations
compared with the maximum likelihood value from the original 
catalogue. Although only a small difference in the ellipticity 
dispersion is brought about by lensing, this is summed over 19400 galaxies
when calculating $\ell$!

Note that changing the value
assumed for the background source redshift, $z$, only has a marginal 
effect on the best-fit NFW parameters: changing $z$ by 0.25 results in
an approximately  5-10\% change in the $M_{200}$ value inferred from
the best-fit values of $r_{200}$ and $c$. This indicates that our
analysis is not too badly affected by the lack of redshifts for the
background sources. However, we hope to obtain photometric redshifts
for many of the background sources in order to obtain as accurate a
result as possible. Assuming an $\Omega=0.3$, $\Lambda=0.7$
cosmology changes the mass estimate by $\approx 5\%$. Again, this
mass estimate is consistent (2.5\% difference) with that of 
Clowe \& Schneider (2001). 

An important point to note is that the difference in the projected mass $M_{\Sigma}$ of A1689 within $R=r_{200}$ for the best-fit SIS, 
power-law and NFW models is less than 20\%. This means that adopting
a particular parameterised model does not lead to a severe error in
the mass estimate.

To conclude, we have applied the shear method discussed in SKE and in
KS to the spectacular lensing cluster Abell 1689. The algorithm
requires only a few seconds of cpu time to obtain best-fit parameterised models. Our results for the best-fit $\sigma_{\rm 1D}$ for the SIS and 
$M_{200}$ for the NFW profile are consistent with standard
$\chi^{2}$ fitting; one main advantage of the maximum likelihood
method is that galaxies need not be binned. 
We are very encouraged by the performance of the method, which will be
applied to a larger sample of clusters in due course.

\begin{acknowledgements}
We would like to thank Simon White for kindly reading the 
manuscript, and an anonymous referee for many helpful comments. 
This work was supported by the TMR Network ``Gravitational Lensing: 
New Constraints on Cosmology and the Distribution of Dark Matter'' 
of the EC under contract No. ERBFMRX-CT97-0172.
\end{acknowledgements}
  
\def\ref#1{\bibitem[1998]{}#1}

\end{document}